\documentstyle[12pt,psfig]{article}
\newcommand{\MeV}{~\mbox{MeV}}
\newcommand{\GeV}{~\mbox{GeV}}
\newcommand{\TeV}{~\mbox{TeV}}
\newcommand{\gsim}{ \mathop{}_{\textstyle \sim}^{\textstyle >} }
\newcommand{\lsim}{ \mathop{}_{\textstyle \sim}^{\textstyle <} }
\newcommand{\vev}[1]{ \left\langle {#1} \right\rangle }
\begin{document}

\baselineskip 1.37em

\setcounter{footnote}{1}

\begin{titlepage}
\begin{flushright}
UT-969\\
hep-ph/0110072\\
\end{flushright}

\vskip 1cm
\begin{center}
{\large \bf Higgsino and Wino Dark Matter from Q-ball Decay\\ in
 Affleck-Dine Baryogenesis}

\vskip 1.2cm 

Masaaki Fujii\,\, and \,\,K.~Hamaguchi
\vskip 0.4cm

{\it Department of Physics\\
University of Tokyo, Tokyo 113-0033, Japan}
\vskip 2cm
\abstract{We claim that the Higgsino-like and wino-like neutralinos can
be good dark matter candidates if they are produced by the late time
decay of Q-ball, which is generally formed in Affleck-Dine baryogenesis. 
The late time decays of the Q-balls into these LSP's and subsequent pair
annihilations of the LSP's naturally lead to the desired mass density of
dark matter. Furthermore, these dark matter can be much more easily
detected by the dark-matter search experiments than the standard
bino-like dark matter.}

\end{center}
\end{titlepage}

\setcounter{footnote}{0}


Supersymmetry (SUSY) has been widely considered as an attractive
framework for physics beyond the standard model. It explains the
stability of the electroweak scale against quadratically divergent
radiative corrections. Furthermore, particle contents of the minimal
SUSY standard model (MSSM) lead to a beautiful unification of the three
gauge coupling constants of the standard model. 

One of the remarkable features in the MSSM is the existence of an ideal
dark matter candidate, that is, the lightest SUSY particle
(LSP).\footnote{We assume that the $R$-parity is exact and hence the LSP
is absolutely stable.} The most extensively studied LSP as a dark matter
candidate is the bino-like neutralino, since its thermal relic abundance
naturally provides a desired amount of the present mass density of the
dark matter.  On the other hand, there have been much less interests in
other candidates, such as the Higgsino-like and wino-like neutralino,
since their thermal relic densities are generally too low to be a
significant component of dark matter~\cite{bino-only}.

In this letter, we claim that the Higgsino-like and wino-like
neutralinos are good dark matter candidates in spite of their large
annihilation cross sections, if the origin of the observed baryon
asymmetry lies in the Affleck-Dine (AD) baryogenesis~\cite{AD,DRT}.  In
AD mechanism, a linear combination of squark and/or slepton fields
($\phi$ field) has a large expectation value along a flat direction
during an inflationary stage, and its subsequent coherent oscillation
creates a large net baryon asymmetry. The coherent oscillation of the
$\phi$ field is generally unstable with spatial perturbation and it
fragments~\cite{Kus-Sha,Enq-McD-1,Enq-McD-2} into non-topological
solitons~\cite{Q-ball}, Q-balls.  It has been, in fact, shown in
detailed numerical calculations~\cite{feq1} that almost all of the
initial baryon asymmetry carried by the $\phi$ field is absorbed into
the Q-balls.

The Q-ball has a long lifetime\footnote{We do not consider the
gauge-mediated SUSY breaking~\cite{GMSB}, where the Q-ball is generally
stable~\cite{Q1over4,Kus-Sha}.} and its decay temperature is likely to
be well below the freeze-out temperature of the LSP, which leads to the
non-thermal production of the dark matter~\cite{Enq-McD-2}.\footnote{In
Ref.~\cite{Enq-McD-2}, the authors only considered the case in which
there is effectively no pair annihilation of the LSP after its
production, and did not consider LSP's with large annihilation cross
sections, such as Higgsino-like and wino-like neutralino.  In general,
their case leads to an overproduction of the LSP. See discussion
below. (See also Refs.~\cite{FHY,McD}.)} We show that in the case of the
Higgsino-like or wino-like LSP the late-time Q-ball decay and subsequent
pair annihilations of the LSP's naturally give rise to a desired
dark-matter energy density.  It is very encouraging that such candidates
are much more easily detected than the standard bino-like neutralino. As
we will see later, the detection rate is in fact more than ten times
larger compared with the case of bino-like LSP~\cite{MW}.

Let us first estimate the present energy density of the non-thermally
produced LSP. Suppose that there is a non-thermal production of LSP at
temperature $T = T_d$, where $T_d$ is below the freeze-out temperature
$T_f$ of the LSP. ($T_f$ is typically given by $T_f\sim m_{\chi}/20$,
where $m_{\chi}$ is the mass of the LSP.) The subsequent evolution of
the number density $n_{\chi}$ of the LSP is described by the following
Boltzmann equation:
\begin{eqnarray}
 \dot{n}_{\chi}
  + 3 H n_{\chi}
  =
  - \vev{\sigma v}
  n_{\chi}^2
  \,,
\end{eqnarray}
where the overdot denotes a derivative with time, $H$ is the Hubble
parameter of the expanding universe, and $\vev{\sigma v}$ is the
thermally averaged annihilation cross section of the LSP. Here, we have
neglected the effect of the pair production of LSP's, which is
suppressed by a Boltzmann factor $\exp(-m_{\chi}/T)$ for $T < m_{\chi}$.
It is useful to rewrite the above equation in terms of temperature $T$
and number density of LSP per comoving volume $Y_{\chi}\equiv n_{\chi} /
s$, where $s = (2\pi^2/45)g_{*} T^3$ is the entropy density and $g_{*}$
is the effective number of relativistic degrees of freedom.  Here and
hereafter, we assume that the energy density of the LSP is much smaller
than that of the radiation $\rho_{rad}$ at $T \simeq T_d$, and no extra
entropy production occurs after that. (We will justify this assumption
later in the case of Q-ball decay.) Then, we obtain\footnote{Here, we
have used $g_{*\rho}\simeq g_*$ for $T\gsim 1\MeV$, where $g_{*\rho}(T)
\equiv( 30/\pi^2) \rho_{rad}(T)/T^4$ is the effective degrees of freedom
for energy density.  }
\begin{eqnarray}
 \frac{d Y_{\chi}}{d T}
  =
  \sqrt{
   \frac{8\pi^2 g_*}{45}
   }
   \left(
    1 + \frac{T}{3g_*}
    \frac{d g_*}{d T}
    \right)
    \vev{\sigma v}
    M_{pl}
    Y_{\chi}^2
    \,,
\end{eqnarray}
where $M_{pl} = 2.4 \times 10^{18} \GeV$ is the reduced Planck scale.
This equation can be analytically solved by using approximations
$g_*(T)\simeq g_*(T_d) \simeq const$ and $\vev{\sigma v}(T) \simeq
const$, which results in
\begin{eqnarray}
 \label{Eq-analytic}
 Y_{\chi}(T)
  \simeq
  \left[
   \frac{1}{Y_{\chi}(T_d)}
   +
  \sqrt{
   \frac{8\pi^2 g_*(T_d)}{45}
   }
   \vev{\sigma v}
    M_{pl}
    (T_d - T)
   \right]^{-1}
   \,.
\end{eqnarray}
Therefore, for sufficiently large initial abundance $Y_{\chi}(T_d)$, the
final abundance $Y_{\chi 0}$ for $T \ll T_d$ is given by
\begin{eqnarray}
 Y_{\chi 0}
  \simeq
  \left[
   \sqrt{
   \frac{8\pi^2 g_*(T_d)}{45}
   }
   \vev{\sigma v}
   M_{pl}
   T_d
   \right]^{-1}
   \,.
\end{eqnarray}
Notice that the final abundance $Y_{\chi 0}$ is determined only by the
temperature $T_d$ and the cross section $\vev{\sigma v}$, independently
of the initial value $Y_{\chi}(T_d)$ as long as $Y_{\chi}(T_d)\gg
Y_{\chi 0}$.  {}From the above formula, we obtain the relic mass density
of the LSP in the present universe:
\begin{eqnarray}
 \Omega_{\chi}
 &\simeq&
  0.5
  \left(\frac{0.7}{h}\right)^2
  \times
  \left(
   \frac{m_{\chi}}{100 \GeV}
   \right)^3
   \left(
    \frac{10^{-3}}{ m_{\chi}^2 \vev{\sigma v}}
    \right)
    \times
    \left(
     \frac{100 \MeV}{T_d}
     \right)
     \left(
      \frac{10}{g_*(T_d)}
      \right)^{1/2}
      \,,
      \label{Omega-ana}
\end{eqnarray}
where $h$ is the present Hubble parameter in units of $100\,\, {\rm
km}\,\, {\rm sec}^{-1} {\rm Mpc}^{-1}$ and $\Omega_{\chi}\equiv
\rho_{\chi}/\rho_c$. ($\rho_{\chi}$ and $\rho_c$ are the energy density
of LSP and the critical energy density in the present universe,
respectively.) Notice that the obtained abundance is much larger than
the result for thermally produced LSP.  In fact, it is enhanced by a
factor of $\sim (T_f/T_d)$ compared with the case of standard thermal
production with the $s$-wave dominant annihilations.

Now let us discuss the LSP production by the Q-ball decay. First of all, 
the baryon number density $n_B$ is related to the number density of the
Q-balls and the initial charge of each Q-ball $Q_i$:
\begin{eqnarray}
 Q_i n_Q
  =
  f n_B
  \,,
\end{eqnarray}
where $f$ denotes the fraction of the total baryon asymmetry which is
initially contained in the Q-balls. Notice that almost all the baryon
asymmetry is initially stored in the Q-balls~\cite{feq1}, namely,
$f\simeq 1$.  

The decay rate of a single Q-ball is given by~\cite{Q-decay}:
\begin{eqnarray}
 \Gamma_Q \equiv - \frac{dQ}{dt} \lsim \frac{\omega^3 {\cal A}}{192\pi^2}
  \,,
\end{eqnarray}
where $\omega\simeq m_{\phi}$, $m_{\phi}$ is the soft scalar mass of the
$\phi$ field, and ${\cal A}$ is the surface area of the Q-ball. (The
radius of the Q-ball is given by $R_Q\simeq
\sqrt{2}|K|^{-1/2}m_{\phi}^{-1}$, where $|K|\simeq
0.01$--$0.1$~\cite{Enq-McD-1,Enq-McD-2,Enq-Jok-McD}.) Then, we obtain
the lifetime of the Q-ball $\tau_d\equiv Q_i/\Gamma_Q$, or equivalently
the decay temperature $T_d$ of the Q-ball:
\begin{eqnarray}
 T_d \lsim 2 \GeV
  \left(
   \frac{0.01}{|K|}
   \right)^{1/2}
   \left(
    \frac{m_{\phi}}{1\TeV}
    \right)^{1/2}
    \left(
     \frac{10^{20}}{Q_i}
     \right)^{1/2}
     \,.
\end{eqnarray}
Actually, the formed Q-ball has a large charge and typically decays at
$T_d\lsim 1\GeV$~\cite{Enq-McD-2}, in particular when the $\phi$ field
is lifted by a nonrenormalizable dimension-six operator in the
superpotential with a cutoff scale $\sim M_{pl}$. Hereafter, we will
take $T_d\simeq 10\MeV$--$1\GeV$. {}From above equations, the production
rate of the LSP per time per volume is given by
\begin{eqnarray}
 N_{\chi} \Gamma_Q n_Q 
  &=&
  N_{\chi} \Gamma_Q
  \frac{f n_B}{Q_i}
   \theta
   \left(
    Q_i - \Gamma_Q t
    \right)
    \nonumber
    \\
 &=&
  N_{\chi}
  f
  n_B
  \times
  \frac{
   \theta( \tau_d - t )
   }
   {\tau_d}
   \,,
\end{eqnarray}
where $N_{\chi}$ is the number of LSP's produced per baryon number,
which is at least $3$. Thus, the evolution of the number density of the
LSP is obtained by solving the following equation:
\begin{eqnarray}
 \dot{n}_{\chi}
  + 3 H n_{\chi}
  =
  N_{\chi}
  f
  \left(
   \frac{n_B}{s}
   \right)_0
   s
  \times
  \frac{
   \theta( \tau_d - t )
   }
   {\tau_d}
  - \vev{\sigma v}
  n_{\chi}^2
  \,,
  \label{B-eq}
\end{eqnarray}
where we have normalized the baryon number density by the entropy
density and subscript $0$ denotes the present value.

In Eq.~(\ref{B-eq}), it is assumed that the LSP is uniformly
distributed. Because the LSP's are produced from the Q-ball, which is a
localized object, one might wonder if the pair annihilation rate of the
LSP becomes much larger and its final number density becomes much
smaller. However, we can see this is not the case as follows. First of
all, it is found from Eq.(\ref{Eq-analytic}) that the number density of
the LSP approaches its final value only after $(T_d - T)/T_d\sim {\cal
O}(1)$, which means it takes a time scale $\Delta t\sim \tau_d$. (Notice
that this is true for any local number density, as long as it is large
enough.) By that time, LSP's have spread out by a random walk colliding
with the background particles, and form a Gaussian distribution around
the decaying Q-ball. The central region of this distribution has a
radius $\bar{r}\simeq \sqrt{\nu \tau_d}$, where $\nu^{-1} \simeq G_F^2
m_{\chi} T_d^4$~\cite{Enq-McD-2}. ($G_F$ is the Fermi coupling
constant.) Meanwhile, we can see that the number of Q-balls within this
radius is much larger than one, roughly given by $(4\pi/3)\,\bar{r}^3 n_Q
\sim
10^{10}\times(T_d/1\!\GeV)^{-6}(Q_i/10^{20})^{-1}(m_{\chi}/100\!\GeV)^{-3/2}$.
Hence, the assumption of the uniform distribution is justified.

We should also note that the energy density of the Q-balls $\rho_Q$ is
much smaller than that of the radiation $\rho_{rad}$ for $T\gsim T_d$:
\begin{eqnarray}
 \frac{\rho_Q}{\rho_{rad}}
  &\simeq&
  \frac{m_{\phi} n_Q Q_i}{\rho_{rad}}
  \nonumber\\
 &=&
  \frac{3 m_{\phi}}{4 T}
  f
  \left(
   \frac{n_B}{s}
   \right)_0
   \nonumber\\
 &\sim&
  10^{-5} f \times
  \left(
   \frac{m_{\phi}}{1 \TeV}
   \right)
   \left(
    \frac{10 \MeV}{T}
    \right)
    \ll 1
    \,,
\end{eqnarray}
where we have used the fact that the energy of the Q-ball per charge is
roughly given by $m_{\phi}$. Therefore, no significant entropy
production takes place during the Q-ball decay.

\begin{figure}[t]
 \centerline{ {\psfig{figure=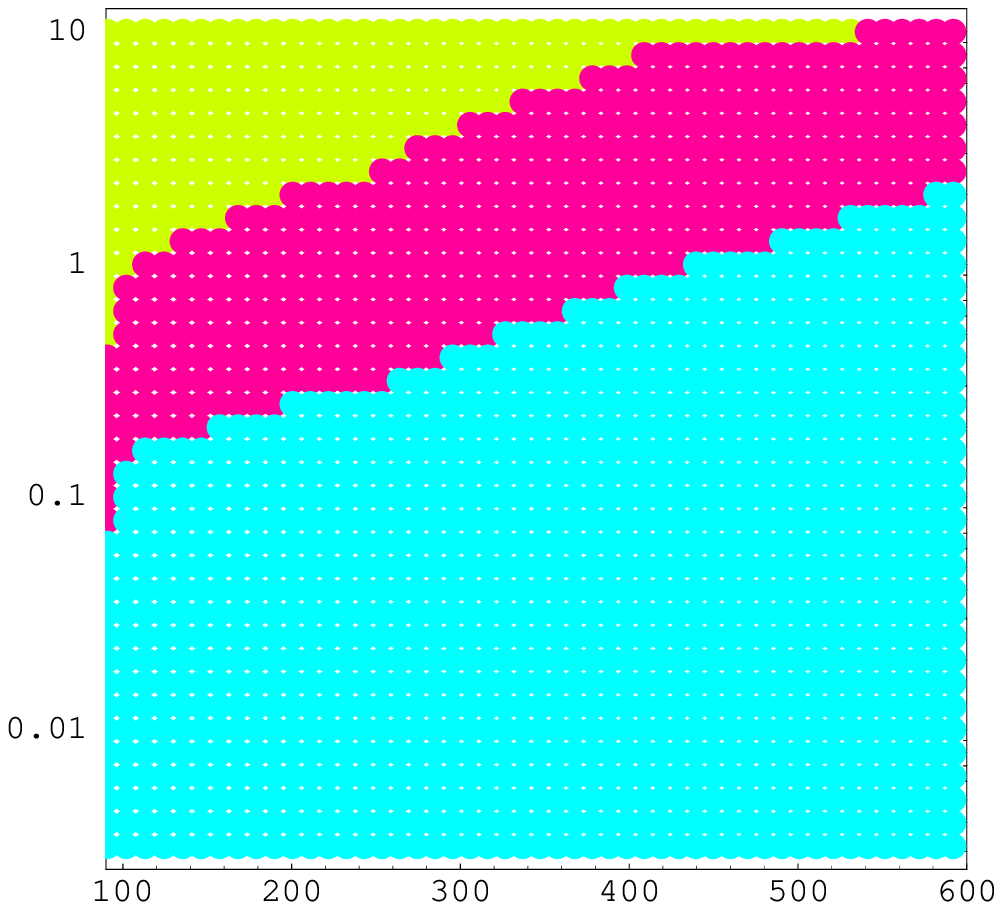,height=7cm}} }
\begin{picture}(0,0)
 \put(70,110){$T_d  \,[\mbox{GeV}]$}
 \put(210,0){$m_{\chi} \,[\mbox{GeV}]$}
\end{picture}
 \caption{The contour plots of the relic abundance $\Omega_{\chi}$ of
 the Higgsino-like neutralino LSP in the $m_{\chi}$--$T_d$ plane. We
 have taken $h=0.7$, $M_1=(3/2)\mu$, $M_2=3\mu$, $\tan\beta = 5$,
 $m_{A^0}=300\GeV$, $a=0$, and $m_0 = 1\TeV$. The three shaded regions
 correspond to the range of $\Omega_{\chi}<0.1$, $0.1<\Omega_{\chi}<1$,
 $1<\Omega_{\chi}$, from the top to the bottom, respectively. }
 \label{Fig-higgsino-sq1000}
\end{figure}

\begin{figure}[h!]
 \centerline{ {\psfig{figure=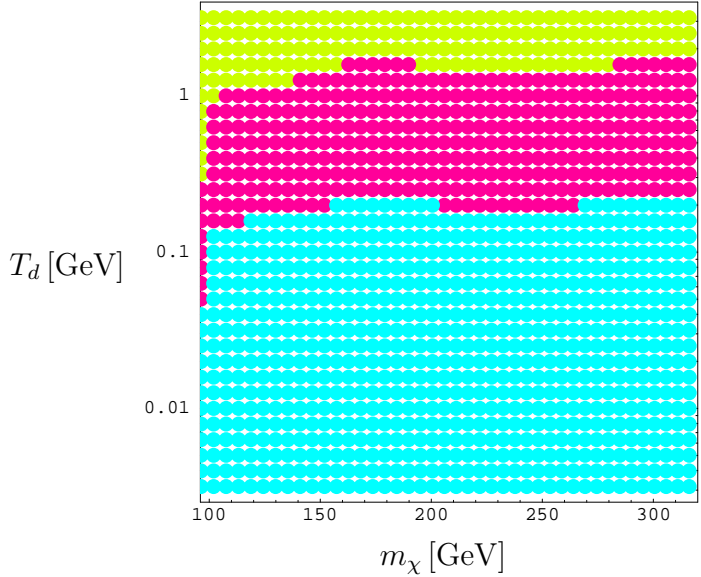,height=7cm}} }
\begin{picture}(0,0)
 \put(70,110){$T_d  \,[\mbox{GeV}]$}
 \put(210,0){$m_{\chi} \,[\mbox{GeV}]$}
\end{picture}
 \caption{The contour plots of the relic abundance $\Omega_{\chi}$ of
 the Higgsino-like neutralino LSP in the $m_{\chi}$--$T_d$ plane. The
 parameters are the same as Fig.~\ref{Fig-higgsino-sq1000} except $m_0 =
 330\GeV$. The three shaded regions correspond to the range of
 $\Omega_{\chi}<0.1$, $0.1<\Omega_{\chi}<1$, $1<\Omega_{\chi}$, from the
 top to the bottom, respectively.}
 \label{Fig-higgsino-sq330}
\end{figure}

\begin{figure}[t]
 \centerline{ {\psfig{figure=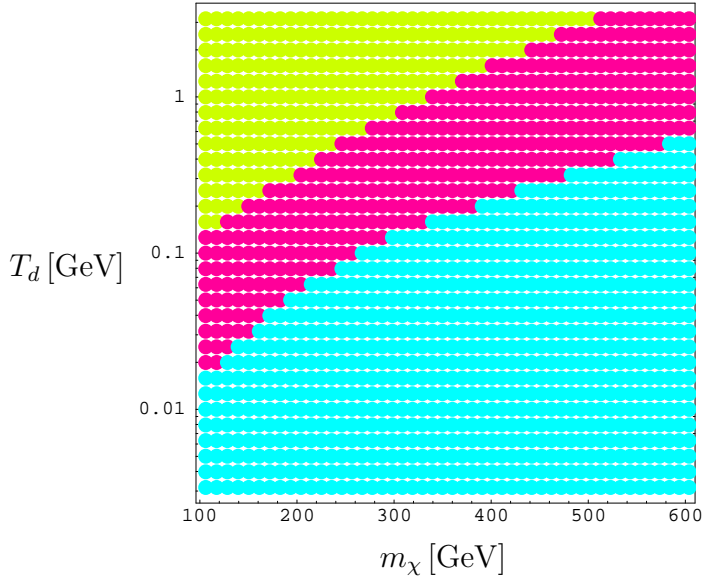,height=7cm}} }
\begin{picture}(0,0)
 \put(70,110){$T_d  \,[\mbox{GeV}]$}
 \put(210,0){$m_{\chi} \,[\mbox{GeV}]$}
\end{picture}
 \caption{The contour plots of the relic abundance $\Omega_{\chi}$ of
 the wino-like neutralino LSP in the $m_{\chi}$--$T_d$ plane. We took
 $M_1 = (3/2)\mu$ and $M_2 = (1/2)\mu$. Other parameters are the same as
 Fig.~\ref{Fig-higgsino-sq1000}. The three shaded regions correspond to
 the range of $\Omega_{\chi}<0.1$, $0.1<\Omega_{\chi}<1$,
 $1<\Omega_{\chi}$, from the top to the bottom, respectively.}
 \label{Fig-wino}
\end{figure}

We have numerically solved the Boltzmann equation (\ref{B-eq}) for
Higgsino-like and wino-like neutralino and calculated the relic
abundance $\Omega_{\chi}$, where we have taken $N_{\chi}=3$, $f=1$,
$(n_B/s)_0=0.7\times 10^{-10}$, and $h=0.7$.  In our calculation, we
included final states; $W^+W^-$, $ZZ$, $t \bar{t}$, $h^0 A^0$, $H^0
A^0$, $Z h^0$, $Z H^0$, and $W^{\pm} H^{\mp}$.\footnote{Here, we
included only $s$-wave annihilation cross sections, which is a
reasonable approximation for Higgsino-like and wino-like neutralino and
for $T\ll m_{\chi}$.}  (We took the cross sections from
Ref.~\cite{Nojiri-cs}.) The results are shown in
Figs.~\ref{Fig-higgsino-sq1000}--\ref{Fig-wino} in the $m_{\chi}$--$T_d$
plane.  Figs.~\ref{Fig-higgsino-sq1000} and \ref{Fig-higgsino-sq330}
correspond to the Higgsino-like LSP, where we took $M_1 = (3/2)\mu$ and
$M_2 = 3\mu$, while Fig.~\ref{Fig-wino} corresponds to the wino-like
LSP, where we took $M_1 = (3/2)\mu$ and $M_2 = (1/2)\mu$. ($M_1$ and
$M_2$ are the soft gaugino masses for the $SU(2)_L$ and $U(1)_Y$ gauge
groups, respectively, and $\mu$ denotes the SUSY contribution to the
Higgs-boson (Higgsino) masses.) As for the other parameters, we used
$\tan\beta = 5$, $m_{A^0} = 300\GeV$ and all trilinear scalar couplings
$a=0$, in all figures. For sfermions we assumed universal soft masses
$m_{\widetilde{f}i}=m_0$. We used $m_0= 1\TeV$ in
Figs.\ref{Fig-higgsino-sq1000} and \ref{Fig-wino}, and $m_0 = 330\GeV$
in Fig.\ref{Fig-higgsino-sq330}.  We used the $g_*(T)$ given in
Ref.~\cite{gstar} for $100\MeV\lsim T\lsim 1\GeV$, adopting the QCD
phase transition near $150\MeV$.

It is found from these figures that both of the the mass densities of
the Higgsino-like and wino-like LSP in fact fall in the desired region
$\Omega_{\chi}\simeq 0.1$--$1$ in a wide range of LSP mass, for Q-ball
decay temperatures $T_d\simeq 10\MeV$--$1\GeV$.\footnote{
In Fig.\ref{Fig-higgsino-sq330}, the annihilation cross section of the
Higgsino is
dominated by the decay mode into $t\bar{t}$ because of the light stop
when $m_{\chi}>m_{t}$. This is why the resultant relic abundance of 
the LSP seems almost constant contrary to the naive expectation.
} (We have also confirmed
that these results are well reproduced by the analytic calculation given
in Eq.~(\ref{Omega-ana}).) It is remarkable that the Higgsino-like and
the wino-like LSP's can be excellent dark matter candidates even in the
relatively small mass region, where the thermal production would give
rise to too small relic abundance. (See discussion below
Eq.(\ref{Omega-ana}).) As we will see later, these regions are also
advantageous for dark matter search experiments.

Here, we should note that the Q-ball decay would produce too large
amount of dark matter density if there were no pair annihilation of the
LSP's. This can be easily seen by integrating the Eq.~(\ref{B-eq}) with
$\vev{\sigma v} = 0$:
\begin{eqnarray}
 \left.
  \Omega_{\chi}
  \right|_{\rm no\,\,ann}
  &=&
  3
  \left(
   \frac{N_{\chi}}{3}
   \right)
   f
   \left(
    \frac{m_{\chi}}{m_n}
    \right)
    \Omega_B
    \nonumber\\
 &\gsim&
  2.6\,f\times
  \left(
   \frac{N_{\chi}}{3}
   \right)
   \left(
    \frac{m_{\chi}}{100\GeV}
    \right)
    \left(
     \frac{0.7}{h}
     \right)^2
     \,,
\end{eqnarray}
where $m_n\simeq 1\GeV$ is the nucleon mass, and we have used the bound
on the present baryon density $\Omega_B h^2\gsim
0.004$~\cite{nBs-bound}. Therefore, in the case of the bino-like LSP,
the Q-ball formation is a serious obstacle for the AD
baryogenesis. (Detailed discussion on this problem and possible
solutions are given in Ref.~\cite{FHY}.)
\\
\\

As we have seen, Higgsino-like and wino-like LSP are promising
candidates for cold dark matter if the AD baryogenesis is responsible
for generating the observed baryon asymmetry in the present universe.
Encouragingly enough, if this is the case, the direct detecting
possibility for these dark matter is enormously enhanced compared with
the case of bino-like dark matter~\cite{MR,MW}.
 
The relevant quantity for direct search experiments is the elastic
neutralino-nucleon scattering rate~\cite{Nojiri-ns}:
\begin{eqnarray}
&&R=\frac{\sigma \rho_{\chi}^{halo}v_{\chi}F_{\xi}}{m_{\chi}M_{N}},
\end{eqnarray}
where $\rho_{\chi}^{halo}$ and $v_{\chi}$ is the mass density and the
average speed of the neutralinos in the galactic halo,
respectively. $M_{N}$ is the mass of the target nucleus, and $F_{\xi}$
is the nuclear form factor.  By using typical values, this scattering
rate is written as
\begin{equation}
R=\frac{\sigma F_{\xi}}{m_{\chi}M_{N}}
\;1.8 \times 10^{11}\GeV^{4}
\;\left(\frac{\rho_{\chi}^{halo}}{0.3\GeV/\mbox{cm}^3}\right)
\left(\frac{v_{\chi}}{320\mbox{km/sec}}\right)
\left(\frac{\mbox{events}}{\mbox{kg}\cdot\mbox{day}}\right).
\end{equation}
The list of relevant coupling constants are given in
Ref.~\cite{Nojiri-ns}.  For the numerical calculations in this work, we
have neglected the squark and Z-boson exchange contributions, since they
are subdominant components in most of the parameter space~\cite{MW}, and 
we have taken $F_{\xi}=1$ for simplicity.

\begin{figure}[h!]
 \centerline{ {\psfig{figure=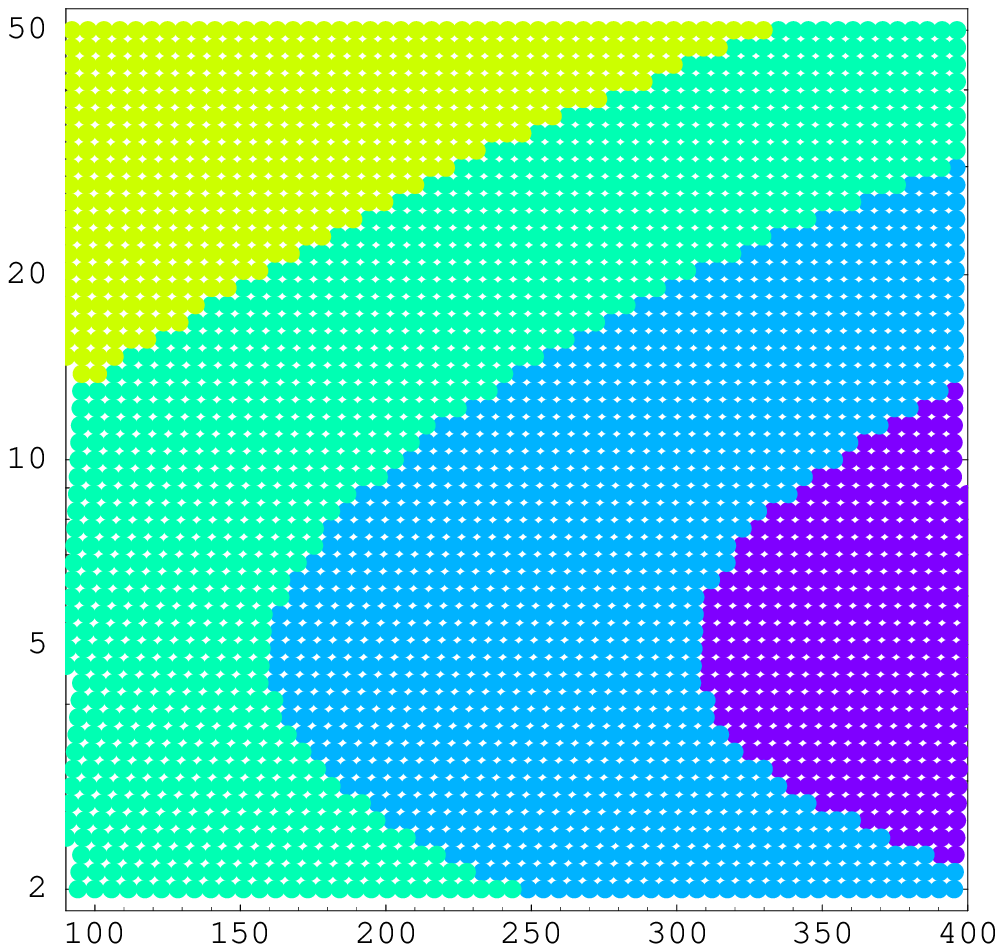,height=7cm}} }
\begin{picture}(0,0)
 \put(80,110){$\tan\beta$}
 \put(210,0){$m_{\chi} \,[\mbox{GeV}]$}
\end{picture}
 \caption{Contours of the detection rate for the Higgsino-like dark
matter in $^{76}Ge$ detector. The four shaded regions correspond to the
ranges of the detection rate $R>0.1$, $0.1\geq R>0.03$, $0.03\geq
R>0.01$, $0.01\geq R\;\;(\mbox{events}/\mbox{kg}\cdot \mbox{day})$ from
left to right, respectively.  The parameters used in this calculation
are the same as in Fig.~\ref{Fig-higgsino-sq1000} except $\tan\beta$.  } 
\label{higgsino-detect}
\end{figure}

\begin{figure}[t]
 \centerline{ {\psfig{figure=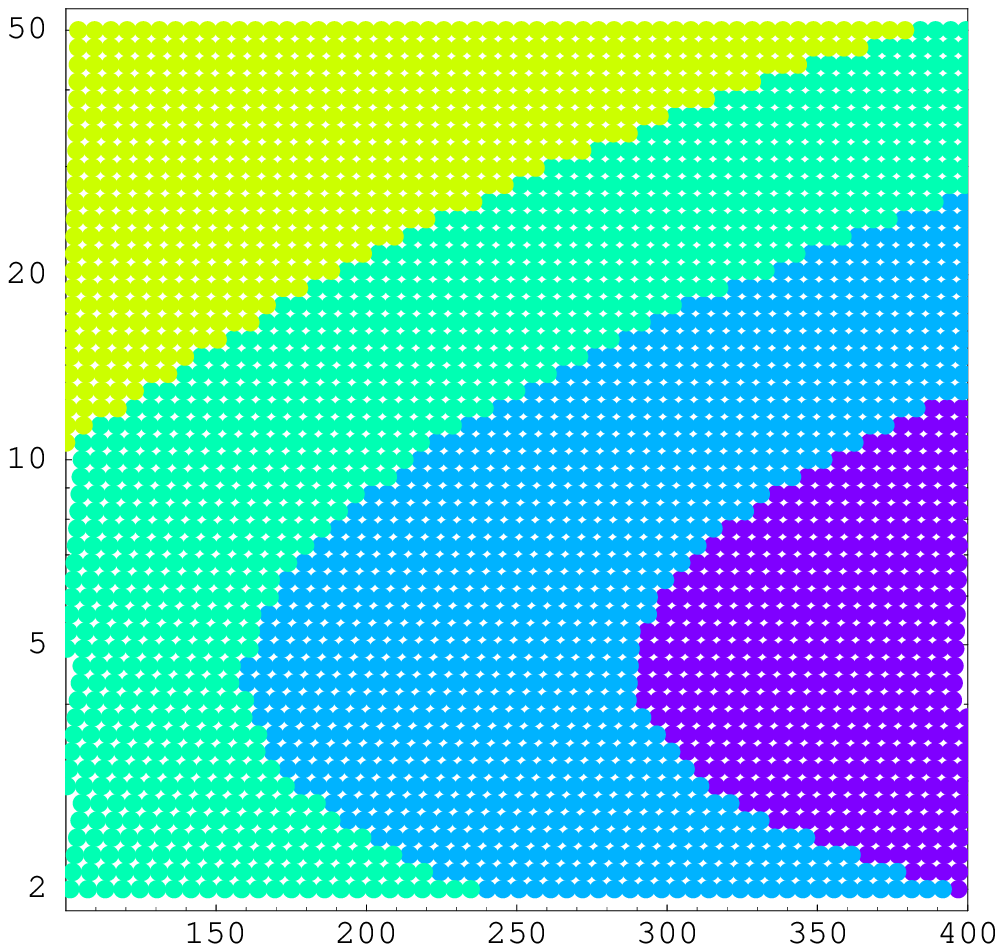,height=7cm}} }
\begin{picture}(0,0)
 \put(80,110){$\tan\beta$}
 \put(210,0){$m_{\chi} \,[\mbox{GeV}]$}
\end{picture}
  \caption{Contours of the detection rate for the wino-like dark matter
in $^{76}Ge$ detector. The four shaded regions correspond to the ranges
of the detection rate $R>0.1$, $0.1\geq R>0.03$, $0.03\geq R>0.01$,
$0.01\geq R\;\;(\mbox{events}/\mbox{kg}\cdot \mbox{day})$ from left to
right, respectively.  The parameters used in this calculation are the
same as in Fig.~\ref{Fig-wino} except $\tan\beta$.}  \label{wino-detect}
\end{figure}

In Fig.~\ref{higgsino-detect} and \ref{wino-detect}, we show the
scattering rates for the Higgsino-like and wino-like dark matter in
$^{76}Ge$ detector, respectively.  In these calculations, we have taken
the same parameter sets as in Fig.~\ref{Fig-higgsino-sq1000} and
\ref{Fig-wino} except $\tan\beta$.  {}From these figures, we see that
the detection rates for the Higgsino-like and wino-like dark matter are
$R\geq 0.01\; \mbox{events}/\mbox{kg}\cdot\mbox{day}$ in very wide range
of parameter space, and they even reach
$R\geq0.1\;\mbox{events}/\mbox{kg}\cdot\mbox{day}$ in the large
$\rm{tan}\beta$ region.  We should stress that such a parameter region
where $R\geq 0.01\; (\mbox{events}/\mbox{kg}\cdot \mbox{day})$ is within
the reach of the on-going cold dark matter searches~\cite{Preport}.

\begin{figure}[h]
 \centerline{ {\psfig{figure=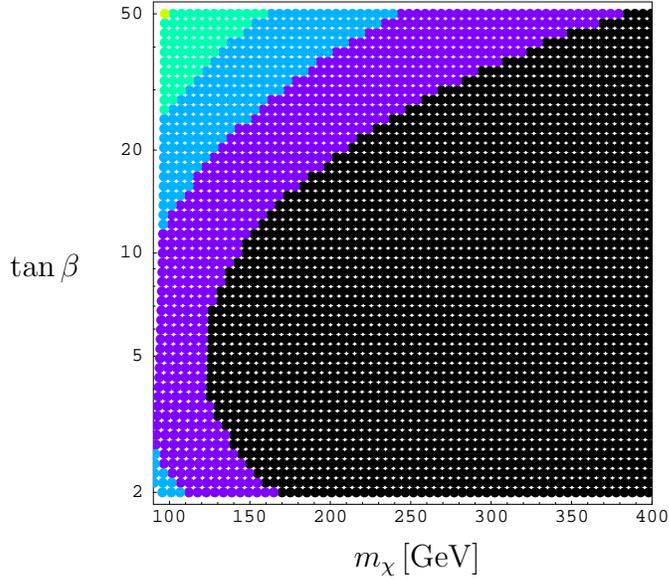,height=7cm}} }
\begin{picture}(0,0)
 \put(80,110){$\tan\beta$}
 \put(210,0){$m_{\chi} \,[\mbox{GeV}]$}
\end{picture}
  \caption{Contours for the detection rate for the bino-like dark matter
in $^{76}Ge$ detector.  The four shaded regions correspond to the ranges
of the detection rate $0.03\geq R>0.01$, $0.01\geq R>0.003$, $0.003\geq
R>0.001$, $0.001\geq R \;\;(\mbox{events}/\mbox{kg}\cdot \mbox{day})$
from left to right, respectively. Here we have taken $M_{1}=(1/3) \mu$
and $M_{2}=(2/3) \mu $.  Other parameters are the same as in
Fig.~\ref{higgsino-detect} and \ref{wino-detect}.}  \label{bino-detect}
\end{figure}

For comparison, we also show the scattering rate in $^{76}Ge$ detector
for the case of bino-like dark matter in Fig.~\ref{bino-detect}.  Here,
we have taken $M_{1}=(1/3) \mu,\;M_{2}= (2/3) \mu$, and other parameters
are the same as in Fig.~\ref{higgsino-detect} and \ref{wino-detect}.  As
we noted, the scattering rate for the bino-like dark matter is much
smaller than Higgsino-like and wino-like dark matter, and the direct
detection in the dark matter search experiments is much more difficult.
\\
\\

To summarize, we pointed out in this letter that the Higgsino-like and
wino-like LSP's can be excellent dark matter candidates if the
Affleck-Dine baryogenesis is responsible for the generation of the
observed baryon asymmetry in the present universe.  Actually, we showed
that the relic abundances of these LSP's can naturally explain the
observed dark matter density with natural Q-ball decay temperatures,
even in the relatively light neutralino mass region, which is much
advantageous for the direct dark matter searches~\cite{Preport}. 

The novel thermal history of the universe proposed in this letter may
have important implications on general SUSY breaking models, which
include mSUGRA, no-scale type models with non-universal gaugino
masses~\cite{noscale-nonU}, anomaly mediated SUSY breaking
model~\cite{A-med}, and so on. Detailed analysis on specific models will
be given elsewhere~\cite{FH}.

In the context of the anomaly-mediated SUSY breaking
(AMSB)~\cite{A-med}, non-thermal production of the wino dark matter from
the late time decay of moduli and gravitino was investigated in
Ref.~\cite{MR}. In the case of moduli decay, however, there exists large
entropy production which substantially dilutes the primordial baryon
asymmetry. The authors suggested that the AD baryogenesis can produce
enough baryon asymmetry even in this case. Although the AD mechanism in
AMSB scenario is generally difficult, there is an attractive AD
scenario~\cite{FHY} which naturally works even in AMSB. However, if
there is a large entropy production from the moduli decay, it is highly
difficult to generate the required baryon asymmetry. Even if this is
possible, the decay of the resultant large Q-ball probably plays a
comparable role with the moduli decay as the non-thermal source of the
LSP's.

\section*{Acknowledgments}
We would like to thank T. Yanagida for various suggestions and
stimulating discussions.  This work was partially supported by the Japan
Society for the Promotion of Science (MF and KH).


\newpage



\end{document}